\documentclass[12pt,preprint]{aastex}

\newcommand{\modot}{M$_\odot$\ }
\newcommand{\msun}{M_\odot\ }
\newcommand{\der}{{\rm d}}
\newcommand{\e}{_{\rm e}}
\newcommand{\lya}{Ly$\alpha$\ }

\newcommand{\fl}{_{\rm floor}}
\newcommand{\cc}{_{\rm c}}
\newcommand{\io}{_{\rm ion}}

\newcommand{\h}{_{\rm h}}
\newcommand{\f}{_{\rm f}}
\newcommand{\m}{_{\rm m}}
\newcommand{\cm}{\, {\rm cm}}
\newcommand{\kms}{\, {\rm km}\, {\rm s}^{-1}}
\newcommand{\X}{_{\rm X}}
\newcommand{\ba}{_{\rm b}}

\newcommand{\kB}{k_{\rm B}}

\newcommand{\SN}{_{\rm SN}}

\newcommand{\diss}{_{\rm diss}}
\newcommand{\fe}{_{e^{-}}}

\newcommand{\shi}{_{\rm shield}}

\newcommand{\first}{^{\rm first}}

\newcommand{\cool}{_{\rm cool}}

\newcommand{\HII}{\hbox{H\hskip1.5pt$\scriptstyle\rm II\ $}}

\newcommand{\HH}{H$_2$\ }

\newcommand{\HHf}{_{{\rm H}_2}}
\newcommand{\etal}{et al.\ }

\begin{document}

\title{The maximum contribution to reionization from metal-free stars}
\author{J. M. Rozas\altaffilmark{1}, J. Miralda-Escud\'e\altaffilmark{2,3}
and E. Salvador-Sol\'e\altaffilmark{1,4}}
\altaffiltext{1}{Departament d'Astronomia i Meteorologia, Universitat de
Barcelona, Mart\'i i Franqu\'es 1, 08028 Barcelona, Spain;
[jrozas,eduard]@am.ub.es}
\altaffiltext{2}{Department of Astronomy, Ohio State University, 140 West
18th Avenue, Columbus, OH 43210}
\altaffiltext{3}{Institut d'Estudis Espacials de Catalunya/ICREA;
miralda@ieec.uab.es}
\altaffiltext{4}{CER on Astrophysics, Particle Physics and Cosmology,
Universitat de Barcelona, Mart\'i i Franqu\'es 1, 08028 Barcelona, Spain}

\begin{abstract}
We estimate the maximum contribution to reionization from the first
generation of massive stars, with zero metallicity, under the
assumption that one of these stars forms with a fixed mass in every
collapsed halo in which metal-free gas is able to cool. We assume that
any halo that has already had stars previously formed in one of their
halo progenitors will form only stars with metals, which are assigned
an emissivity of ionizing radiation equal to that determined at $z=4$
from the measured intensity of the ionizing background. We examine the
impact of molecular hydrogen photodissociation (which tends to reduce
cooling when a photodissociating background is produced by the first
stars) and X-ray photoheating (which heats the atomic medium, raising
the entropy of the gas before it collapses into halos).
%These effects limit the number of first generation stars that can be
%formed at high redshift.
We find that in the CDM$\Lambda$ model
supported by present observations, and even assuming no negative
feedbacks for the formation of metal-free stars, a reionized mass
fraction of 50\% is not reached until redshift $z\sim 12$. The
combination of ordinary metal-enriched stars and early metal-free stars
can yield a CMB optical depth to electron scattering of at most $0.13$.
The contribution of metal-free stars to the present Cosmic Infrared
Background is negligibly small.

\end{abstract}

\keywords{cosmology: theory --- diffuse radiation --- intergalactic medium
--- galaxies: formation}

\section{Introduction}

  In the last few years, our understanding of the reionization history
of the universe has greatly improved thanks to the detection of quasars
at increasingly high redshift (e.g., Becker et al.~2001;
Fan et al.~2002; Hu et al.~2002; White et al.~2003) and the new data
on the Cosmic Microwave Background (CMB) radiation from WMAP (Bennet et
al.~2003). Observations from high redshift quasars indicate a fast
increase of the intensity of the ionizing background with cosmic time
occurring at $z\simeq 6$, probably signalling the end of reionization
(Fan et al.~2002), or the epoch when the low-density intergalactic
medium (hereafter, IGM) became fully ionized. Reionization is expected
to have occurred over an extended period of time, during which sources
gradually reionized every atom in the IGM (e.g., Gnedin 2000). Indeed,
the detection of \lya emission from sources at $z>6$ (Hu et al.~2002;
Kodaira et al.~2003; Cuby et al.~2003; Kneib et al.~2004; Rhoads et
al.~2004) that are not highly luminous to have produced large \HII
regions around them implies that the IGM could not be fully neutral at
the redshift of these sources, otherwise such photons would have been
scattered out of the line of sight (Miralda-Escud\'e \& Rees 1998;
Madau \& Rees 2000).

  Another important probe to the epoch of reionization is the optical
depth to electron scattering of CMB photons, $\tau\e$. The first
measurement of $\tau\e$ was reported by the {\it Wilkinson Microwave
Anisotropy Probe} mission; although its value is still rather
uncertain (Kogut et al.~2003; Spergel et al.~2003), the measurement
favored an early start to reionization, such that the fractional
ionization of the IGM would have reached 50\% at $z\sim 15$. It is
worth noting that this does not by itself contradict the appearance of
the Gunn-Peterson trough (which marks the end of reionization) at
$z\simeq 6$, because reionization may advance gradually over a long
period of time. However, an early start of reionization presents a
problem in the Cold Dark Matter model of structure formation, in which
only a small fraction of matter has collapsed into halos that can form
stars at $z> 15$, and therefore one needs to assume a very high rate
of emission of ionizing photons from this first generation of stars
(e.g., Haiman \& Holder 2003; Chiu, Fan, \& Ostriker 2003; Onken \&
Miralda-Escud\'e 2004). One possibility is that the first collapsed
halos in which gas was able to cool were efficient in driving gas to
accrete onto central black holes that were formed by the core collapse
of the first massive stars. However, as argued by Dijkstra et
al.~(2004a), any such population of high-$z$ Active Galactic Nuclei
(AGN) would likely contribute a smooth component to the present X-ray
background at energies near $\sim 1$ keV that may be higher than is
allowed by observational constraints (although significant calibration
uncertainties of the X-ray background intensity remain which could
increase the upper limit on the fraction of the X-ray background in a
smooth component). The other possibility is that the first massive
stars, which were metal-free and highly efficient at producing
ionizing photons, were responsible for an early reionization (Oh \etal
2001, Yoshida \etal 2003, Sokasian \etal 2004).

  Much work has been done to characterize the properties of these
metal-free stars (Abel \etal 1998, 2002;
Bromm, Coppi, \& Larson 1999, 2002; Oh \etal 2001).
The formation of the first stars was governed by
molecular hydrogen radiative cooling, which becomes effective at
gas temperatures above $\sim 1000$ K (Yoshida et al.~2003). This virial
temperature is first reached in halos of total mass $\sim 10^6$ \modot,
which start becoming substantially abundant at $z \approx 20$.
Metal-free stars have higher effective surface temperatures than their
metal-rich counterparts during their main-sequence phase. In addition,
they are nearly fully convective and can burn most of their hydrogen
content during their main-sequence lifetime. This makes them highly
efficient as producers of ionizing radiation (emitting $\approx 10^5$
ionizing photons per baryon; Schaerer 2002), most of which can escape
into the IGM due to the small amount of gas in the host halos.

  Despite these advantages of metal-free stars as sources of ionizing
radiation, the idea that these stars may have been responsible for an
early reionization faces a number of difficulties. First, it is
unlikely that many massive stars might form simultaneously in the
first halos where star formation took place, which have low velocity
dispersion and can therefore easily lose their gas after it is heated
by photoionization and supernova explosions. Numerical simulations
suggest that one central massive star will initially form, with a mass
of $\sim 200$ \modot (e.g., Abel et al.~2002; Bromm et al.~2002),
containing a fraction of only $\sim 10^{-3}$ of the baryonic mass of
the halo from which it formed. The ionization and the supernova
explosion resulting from this star can then eject the rest of the gas
from the halo (Bromm, Yoshida, \& Hernquist 2003; Whalen, Abel, \&
Norman 2004, Kitayama \etal 2004), yielding a very low efficiency to
form stars. Later, when the gas falls back into a halo with a total
mass increased by the merging process, it will already be enriched by
metals and will form the next generation of stars with properties that
are presumably similar to present stellar populations in normal
galaxies. If the metal-free star releases about $\sim 100 \msun$ of
metals, a total mass of $10^7 \msun$ of baryons may be polluted to a
metallicity of $10^{-3.5} \msun$, above which cooling is already
modified by the presence of metals, while the emission properties of
massive stars are also modified for even lower metallicities. In cases
where a metal-free star collapses into a black hole releasing no
metals, any cold gas reaching the halo center at a later time will
also not form another metal-free star because the black hole will be
in the center.

  The second difficulty is that the radiation emitted by these first
stars can act as a negative feedback mechanism to slow down the
formation of other metal-free stars in new collapsed halos with pristine
gas. A widely studied effect is photodissociation of molecular hydrogen
(Haiman, Abel, \& Rees 2000). Another feedback effect is the heating of
the IGM by X-rays emitted by supernovae or by the metal-free
main-sequence stars themselves (Oh \& Haiman 2003), which raises the
entropy of gas accreting into halos, resulting in a more stringent
requirement for the gas to cool. This effect has not
been so widely studied as the molecular photodissociation one, but as we
shall see below it may be even more important.

  The aim of this paper is to determine the maximum possible contribution to
reionization from zero metallicity stars, under the assumption that only one
such star forms in each halo, with a fixed stellar mass. Special attention will
be paid to the way that the radiative feedback mechanisms of photodissociation
and X-ray heating limit the formation rate of these metal-free stars. We shall
use a semianalytic model based on merger trees (similar to the approach of,
e.g., Haiman et al.~2000 and Wise \& Abel 2005), including a new ingredient to
compute the halo age distribution and derive the fraction of halos that contain
a metal-free main-sequence star at any given time.

  The paper is organized as follows. In \S~\ref{smethod}, we present our method
to calculate the formation rate of metal-free stars, the photodissociating and
X-ray background they create, and how these radiation backgrounds affect the
formation rate by altering the entropy profile in collapsed halos and the
molecular hydrogen abundance. The results are presented in \S~\ref{res},
showing how each of the two feedback effects we consider impact the rate at
which metal-free stars can form and their contribution to reionization.
A discussion of these results follows in \S~\ref{sdisc}.

\section{Method}
\label{smethod}

  Our central assumption in this paper is that when a halo that contains
metal-free gas (i.e., a halo for which no star has previously formed in
any of its merger-tree progenitors) collapses,
a single massive star forms at the halo center after a cooling time.
The mass of this star is assumed to be constant for all halos with
metal-free gas. As explained in detail later (\S~\ref{sstform}), the star forms
in our model if the cooling time of the gas at the center of the halo is
shorter than the lifetime of the halo. We develop an algorithm to define
the lifetime of halos from the merger tree framework based on the
extended Press-Schechter model, which is fully described in the appendix.
The cooling time depends on the central gas density and temperature,
which depends at the same time on the initial entropy of the gas prior to
the collapse of the halo (e.g., Oh \& Haiman 2003). We compute the
initial central gas density (prior to the onset of cooling) in halos at
each redshift self-consistently, assuming hydrostatic equilibrium in a
halo with the Navarro, Frenk, \& White (1997) density profile for the
dark matter (hereafter, NFW).
% and computing the initial entropy as a
%function of redshift as affected by heating from the X-ray background
%generated by the metal-free stars and their supernova explosions.
We calculate the ionizing radiation emitted by the metal-free stars to
infer the history of reionization (taking into account an escape
fraction of ionizing photons from halos determined by recombinations in
the halo
gas), as well as the feedback effects of X-ray heating (and consequent
entropy raise of the atomic medium, which affects the gas density
profiles in halos) and molecular photodissociation. We also include a
component of ionizing radiation from a normal stellar population assumed
to form in metal-enriched halos, which have had previous star formation
and ionization. This section explains in detail how all these
ingredients are implemented in our model. Our goal will be to find the
maximum contribution that these metal-free stars could plausibly make
to an early reionization, and how this is limited by the minimal
expected feedback effects.

\subsection{Mass and Age Distribution of Halos}

  We use the halo mass function of Sheth \& Tormen (2002). We assume
that a first-generation star forms when the halo age exceeds the cooling
time at the halo center. We adopt for this purpose the definition of
halo formation time of Lacey \& Cole (1993), which is the time when the
halo mass was half the present one. The probability distribution of the
age of a halo as a function of its mass and redshift is calculated
following Lacey \& Cole, which we then use to compute the probability
that a halo contains a metal-free star as described in the Appendix.

\subsection{Gas Density Profile}\label{density}

  To determine the cooling time at the halo center, we assume the gas is
in hydrostatic equilibrium in the halo gravitational potential, and that
the dark matter follows a NFW density profile with concentration
parameter $c=5$. In reality, $c$ depends on the halo mass and redshift
(Bullock et al.~2001; Eke et al.~2001), but here we approximate it as
constant and with the same value used by Oh \& Haiman~(2003).

  Under the assumption of hydrostatic equilibrium, the gas density
profile depends on its temperature. Shock-heating of the gas during
the collapse heats the gas to a roughly constant temperature, about
equal to the halo virial temperature. This results in a final gas
density profile similar to the dark matter, although with a small,
constant density core, with a size that is determined by the minimum
entropy $K_{min}$ of the gas in the halo. The gas with the lowest
entropy is generally the one that is shock-heated to the virial
temperature only after reaching a high density, and ends up near
the halo center. However, if the gas had already been heated prior to
the collapse, acquiring a high initial entropy, the value of $K_{min}$
may be increased, resulting in a much more extended gas profile
and reduced central gas density, and therefore a much longer central
cooling time. To take this into account we follow the procedure
explained in Oh \& Haiman~(2003). We define the function $K$ as
\begin{equation}
  K={T\over n_g^{\gamma-1} } ~, 
\end{equation}
where $\gamma$ is the adiabatic index (equal to $5/3$ for monatomic
gas), and $T$ and $n_g$ the gas temperature and number density. Note
that the entropy of a gas is actually the logarithm of $K$, although
in this paper we refer to the function $K$ as ``entropy'' for
convenience. The entropy before the collapse is what we call the entropy
floor, $K\fl(t)$, which we calculate as a function of redshift as
described in \S \ref{xr}.

\begin{figure}
\plotone{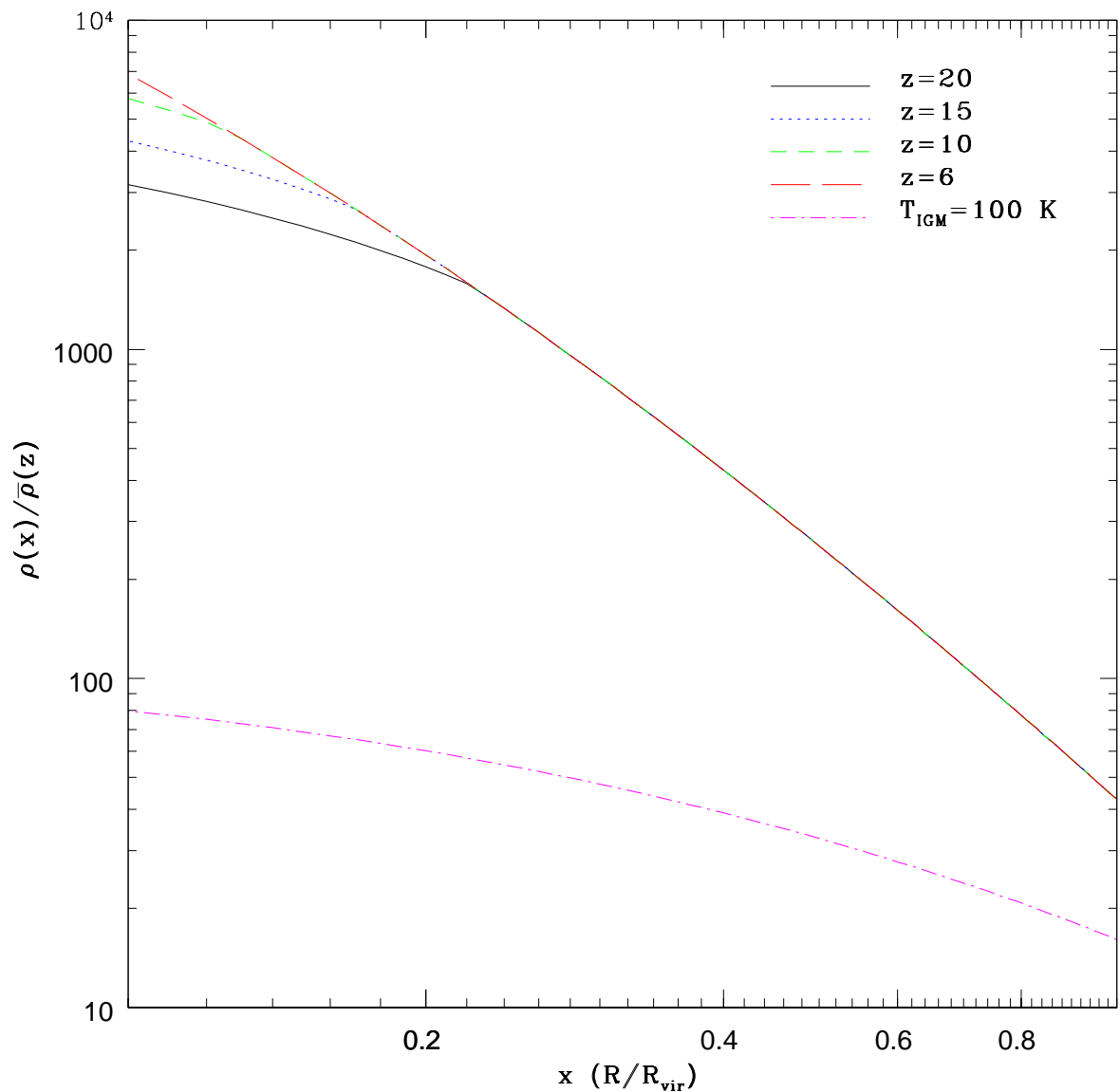}
\caption{Density profiles for a halo of mass $M=5\times 10^5$ \modot.
Different lines correspond to halos at different redshift: $z=20$
(long dashed line), 15 (short dashed line), 10 (dotted line), and 6
(solid line) with primordial entropy. The last case (dotted dashed line)
correspond to redshift 10 but with $T_{IGM}=100~K$.}
\label{fprof}
\end{figure}

  As an example, we show in Figure \ref{fprof} the gas density profile
in a halo with $M=5\times 10^5$ \modot formed at different redshifts.
The initial entropy is fixed to the primordial value given in Oh \&
Haiman (2003), determined by early Compton heating by the CMB on the
adiabatically cooling gas, with the residual ionized fraction left over
from recombination. At $z < 20$, the entropy of the IGM
is practically constant and the entropy floor decreases in importance
with decreasing redshift because of the decreasing gas density, hence
the gas density profile becomes steeper at low redshift. We show also
the effect of raising the IGM temperature to $100$ K at $z=10$, which
leads to a large reduction in the central gas density.

\subsection{Cooling}
\label{scool}

  We define the cooling timescale as the ratio of the internal energy
of the gas to the cooling rate per unit of volume, $L$:
\begin{equation}
t\cool=\frac{1.5n_g\kB T}{L} ~,
\label{tcool}
\end{equation}
where $\kB$ is the Boltzmann constant.
We include only molecular hydrogen cooling for metal-free gas. In
practice, atomic cooling can be neglected because metal-free stars form
very rarely in halos with virial temperature above $\sim 8000$ K in our
model. The reason is that these halos usually have some progenitor of
lower virial temperature that has already formed a star previously (this
will be seen in Fig.\ 2). In this case, we consider that the gas is
already enriched with metals and that the emissivity per unit mass is
constant, as described in \S \ref{semis_met_rich}.

  The cooling rate (due to collisional excitation of molecular hydrogen
roto-vibrational lines) is $L_{\HHf}= n_g^2(r)\, f\HHf\, \Lambda\HHf$,
where $f\HHf$ is the molecular fraction. We use the cooling function
$\Lambda\HHf$ of Galli \& Palla~(1998). The molecular fraction $f\HHf$
is determined by the continuous formation of molecules in the gas from
$H^{-}$, which is at the same time made by the primordial concentration
of electrons from the recombination epoch (see Tegmark et al.\ 1997).
We use equation (16) of Tegmark et al., although simplified and modified
to consider \HH dissociation in the following way:
\begin{equation}
\dot f\HHf(r,t) = k_2 n_g f\fe(r,t)-k\diss ~ ,
\end{equation}
where $k\diss$
is the dissociation rate (which we discuss later in \S~\ref{sdiss})
and $f\fe(r,t)$ is the electron fraction, given by equation (\ref{xr:fe}).
The constant $k_2$ determines the formation rate of $H^{-}$ (which
determines the formation rate of molecules), and is given by
$k_2(T)= 1.83\times 10^{-18}T^{0.88}$ cm$^3$ s$^{-1}$. The initial,
primordial value of $f\HHf$ is taken as $2\times 10^{-6}$
(Galli \& Palla~1998). Note that the fraction of hydrogen that forms molecules
from $H^{-}$ within collapsed halos is much larger than this primordial value,
and therefore
the long-term evolution of the molecular hydrogen fraction does not
strongly depend on its initial value.

\subsection{Star Formation}
\label{sstform}

  As mentioned earlier, the main assumption in this work is that only
one metal-free star is formed in each halo that collapses from
progenitors that have never formed a star previously. The mass of the
star, $M_\star$, is much lower than the baryonic mass of the halo, and is in
fact lower than the gas mass in the core of the gas distribution.
Consequently, we can evaluate the cooling time at the halo center and
consider that the star will form when the age of the halo is equal to
this cooling time. The star forms in a halo of mass $M$ with formation
time $t_f$ at a time $t_1$ that obeys
\begin{equation}
  t_1 = t_f + t\cool(M,t_f) ~.
\label{cool_c1}
\end{equation}
The star will then live up to a time
\begin{equation}
 t_2=t_1+ t_{mf}(M_*) ~,
\label{cool_c2}
\end{equation}
where $t_{mf}(M_*)$ is the main-sequence lifetime of a metal-free star
with mass $M_*$.
The probability that a halo of mass $M$ at time $t$ has already formed
a star can be evaluated by computing first numerically the formation
time $t_{f1}$ that obeys equation (\ref{cool_c1}) with $t_1=t$, and then
calculating the probability $P_f(M,t,t_f < t_{f1})$ that the halo
formation time is earlier than $t_{f1}$ (see eq.\ [2.26] of Lacey \&
Cole 1993). Similarly, the probability that
the halo contains a metal-free star on the main-sequence at time $t$ is
evaluated as the probability $P_f(M, t, t_{f2} < t_f < t_{f1})$, where
$t_{f2}$ is the solution of $t_f$ of equation (\ref{cool_c2}) when
$t_2=t$.

  This condition, however, would be valid only if the halo gas were
always metal-free. In reality, if a star had already formed in one of
the halo progenitors, the supernova explosion from that star would
already have polluted the gas with metals and blown it away. Later on,
as the sequence of halo mergers continues, the final halo present at
time $t$ will accrete this gas, but since the gas is metal enriched it
will form metal-rich stars. The emission from metal-rich stars will be
treated differently in our model, as we describe later in \S 
\ref{semis_met_rich}. Here, our main assumption is that the very
massive stars we consider that are highly efficient at emitting ionizing
radiation are formed only when the gas is totally free of heavy elements.

  To take this into account, we evaluate at each time $t$ the average
number of metal-free stars that have formed in a halo of mass $M$, as
(see the Appendix)
\begin{equation}
N_*(M,t)=\int_{0}^t dt'\int^M_0 \der M' N_{LC}(M',t'\rightarrow M,t)
\frac{ dP_*(M',t')}{dt'}\, ,
\end{equation}
where $N_{LC}(M',t'\rightarrow M,t) \, dM'$ is the number of halos of
mass between $M'$ and $M'+dM'$ at time $t'$ that have been incorporated
into a halo of mass $M$ at time $t$ (see eq.\ [2.25] in Lacey \& Cole
1993), and $dP_*(M',t')/dt'$ is the probability per unit time that a
star forms at $t'$ in a halo of mass $M'$. The way the probability
$dp_\star/dt'$ is calculated is described in the Appendix. We then
assume that there is a Poisson distribution of the number of stars that
have formed within any halo, so that the probability that none has
formed is $\exp(-N_*)$. Therefore, the probability that a halo of mass
$M$ harbours a main-sequence star at time $t$ is
\begin{equation}
P_{ms}(M,t) = \int_{t_{f2}}^{t_{f1}} dt' \frac{d P_f(M,t,t')}{dt'}
 \times e^{-N_*(M,t')} ~ ,
\end{equation}
where $\frac{d P_f(M,t,t_f)}{dt_f}$ is the differential probability that a
halo at time $t$ was formed at time $t_f$ and is given by equation
(2.19) in Lacey \& Cole (1994). In practice, the lifetime of the
main-sequence star ($\sim 3\times 10^6$ years) is short enough that
the above integral can be approximated as
\begin{equation}
P_{ms}(M,t) \simeq \frac{dP_f(M,t,t_{f1})}{dt_f}\, t_{mf}(M_*) \times
e^{-N_*(M,t_{f1})} \, .
\label{prob_sh_star}
\end{equation}
%metal-free star has
%formed in a halo of mass $M$ by the time $t$, without having formed in
%any of its previous progenitors, is
%\begin{equation}
%P_*(M,t)=P_f(M, t, t\f<t_{f1}) \times e^{-N_*(M,t)}\, .
%\label{prob_star}
%\end{equation}

\begin{figure}
\plotone{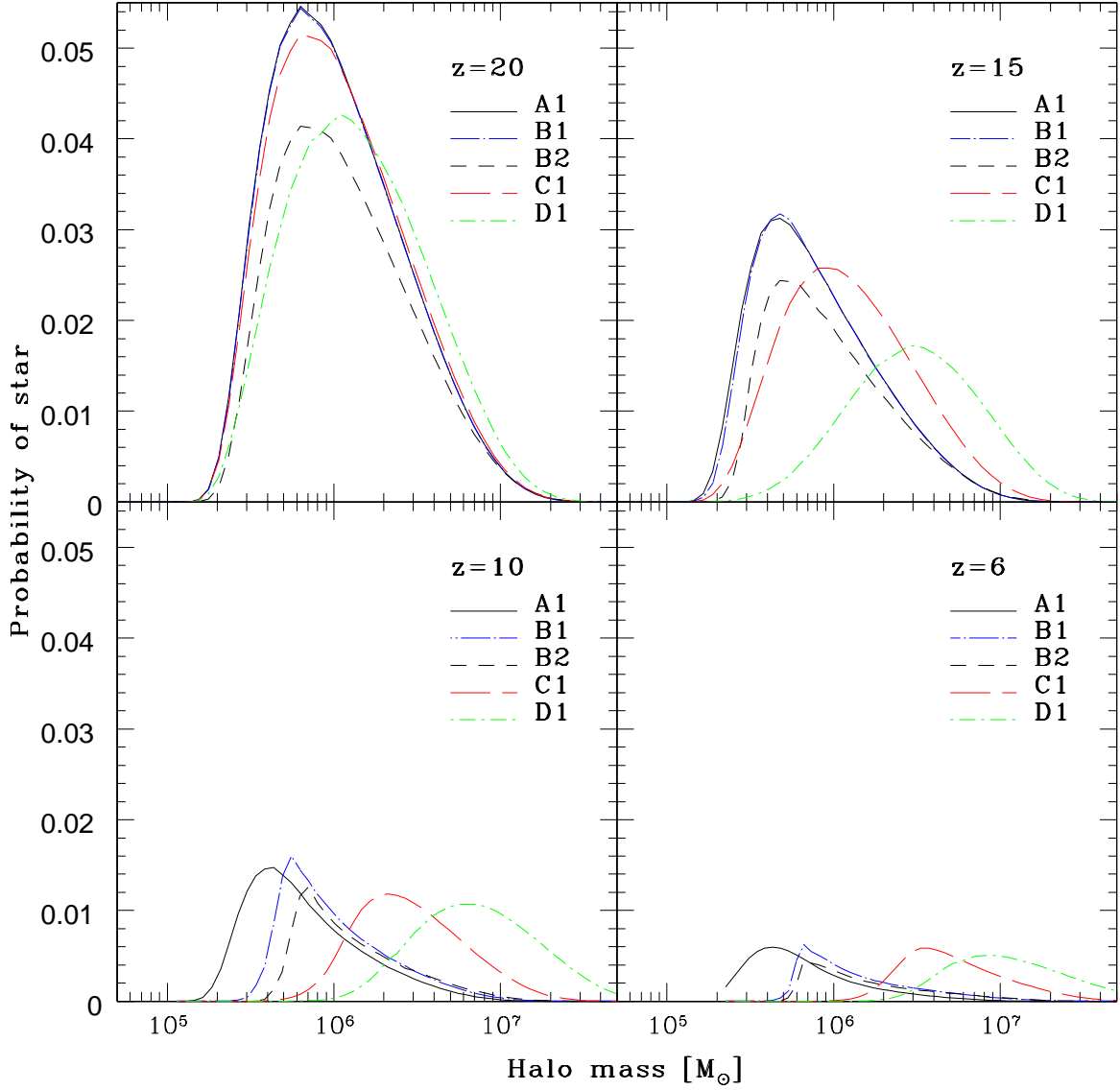}
\caption{ Probability that a hydrogen-burning metal-free star is present in a
halo at a certain redshift as a function of halo mass. The different cases are
those referred in Table \ref{tsub_cases}}
\label{fnstarm}
\end{figure}

  Figure \ref{fnstarm} shows the probability that a halo with mass $M$
harbours a metal-free main-sequence star. Different plots correspond
to different redshifts ($z=20$, 15, 10, and 6), and different lines to
different models that will be described later. The model A1 corresponds
to no feedback effects on the formation of metal-free stars, and the
other models incorporate feedback effects that are described in \S
2.5. The probability to contain a star tends to peak at halo masses of
several times $10^5 \msun$. At lower masses the molecular cooling rate
is too slow and stars have not yet formed, while in halos of higher
mass a metal-free star has typically already formed in one of their
lower-mass progenitors (i.e., the second term in eq.\
[\ref{prob_sh_star}] is very small).

As a byproduct, we can also calculate the global formation rate of
metal-free stars (see the Appendix) as
\begin{equation}
\dot N_\star(t)=\int^\infty_0 \der M\,N\h(M,t)
\,\frac{\der P_\star[M,t]}{\der t}\,.
\label{dndt}
\end{equation}
where $\dot N_\star(t)$ is the metal-free star formation rate and
$N\h(M,t)$ is the Sheth-Tormen halo mass function. This last formula
is useful to calculate the preheating due to X-rays (see section
\ref{xr}).

\subsection{Feedback Effects}

  We consider three feedback processes that affect metal-free star formation.
The first is halo photoevaporation due to the ionizing background (section
\S~\ref{sevap}). The second, photodissociation of molecules due to a soft
ultraviolet background (section \S~\ref{sdiss}). Finally, the third is
reheating of the IGM caused by X-rays (section \S~\ref{xr}).

\subsubsection{Halo Ionization}
\label{sevap}

  The ionizing background reheats the gas in halos and causes a large
increase in the entropy floor. Gas trapped in minihalos escapes
(e.g., Shapiro et al.\ 2004), quenching star formation. Cooling and star
formation in reionized regions can only occur in halos that are massive
enough for the gas to undergo dissipation by atomic cooling. We neglect
any possible formation of metal-free stars in these halos. In practice,
once these halos collapse they should almost always have a progenitor
that already formed a metal-free star in the past, because the number
of halo progenitors increases with halo mass; moreover, even if one
metal-free star may form in one of these more massive halos cooling
mainly by atomic processes, their contribution to the total number of
metal-free stars formed should be negligible because the abundance of
high-mass halos is greatly reduced compared to halos of low mass. We
therefore neglect any possible formation of metal-free stars in
reionized regions. To take this into account, equation (\ref{prob_sh_star})
is multiplied by a factor $1-f\io(t)$, equal to the fraction of the
IGM that is neutral at every redshift. The calculation of the ionized
fraction of the IGM as a function of time using the global rate of star
formation that we compute is described below in \S 2.6. Actually,
the factor $1-f\io$ represents a lower
limit to the suppression of the star formation by reionization, because
in practice halos are correlated, or biased relative to the mass, so
that more halos should be present in the reionized regions where other
high-mass halos have already formed.

\subsubsection{Molecular Photodissociation}
\label{sdiss}

\HH is easily photodissociated by soft ultra-violet photons
(Haiman et al.~2000). The rate at which dissociation is produced can be
approximated by (Abel et al. 1997)
\begin{equation}
k\diss = 1.38\times10^{-12}\, J_{21}(h\nu=12.87 eV)\, F\shi\,
{\rm s}^{-1} ~,
\label{eq:kdiss}
\end{equation}
being $J_{21}(h\nu)$ the flux at frequency $\nu$ in units of
$10^{-21}\,\rm{erg} \rm{s}^{-1} \rm{cm}^{-2} \rm{Hz}^{-1} \rm{str}^{-1}$ and
$F\shi$ the self-shielding factor given by Draine \& Bertoldi~(1996).

The mean free path of dissociating photons in the IGM is typically
larger than the mean distance between minihalos. Therefore, this flux
can be approximated as homogeneous and isotropic. To calculate this
background we apply equation (7) of Haiman et al.\ (2000). The
emissivity of dissociating photons ($j_\nu$ in their formula) is
computed similarly to that of ionizing ones (see \S \ref{semis_met_free}),
although we take into account only the emission by the metal-free stars.
Emission from the metal-rich stars has a small effect at the high
redshifts of interest, which we have neglected.

\subsubsection{Reheating of the IGM by X-rays}\label{xr}

  As proposed by Oh \& Haiman (2003), soft X-ray photons raise the entropy
floor, greatly reducing cooling and star formation in low-mass halos.
X-rays can have a long mean free path through the atomic IGM and can
therefore heat this medium in an approximately uniform way, whereas
the more abundant ultraviolet ionizing photons heat the medium only
near the boundaries of HII regions. The mean free path in the atomic
IGM is roughly
\begin{equation}
l_\nu(z)=\frac{42500}{(1+z)^3} \left(\frac{h\nu}{1 \rm kev}\right)^3 \,
{\rm Mpc} \,.
\end{equation}
When this mean free path is in the range that goes from the mean
distance between neighboring halos containing first-generation stars
(which is about $\sim 100$ kpc at $z=15$; see Yoshida et al.\ 2003) to
the horizon radius ($\sim 70$ Mpc at $z=15$), the photons are effective
at heating the atomic IGM. This corresponds to a range of frequencies
from 200 eV to 2 keV, or the soft X-ray band.

%Ionizations produced by X-rays lead to very energetic
%electrons. This energy is thermalized first with other charged particles
%through Coulomb collisions and later on with neutral atoms by means of
%collisions involving dipole interactions. X-rays can in this way produce a
%raise in the entropy floor of the IGM, which leads to hotter halos, with
%flatter density profiles. Ultraviolet photons, on the contrary, do not produce
%a significant reheating. They have a much shorter mean free path and most of
%the energy is radiated away in recombinations. To see this, we are going to

  Soft X-ray photons can be produced by metal-free stars during two
phases. First, these stars have a high enough effective temperature
during the main sequence lifetime to emit substantially in soft X-rays
(Schaerer 2002). Second, X-rays can be emitted by the hot gas in a
supernova remnant after the death of the star. For simplicity we
consider that all the emission is produced at the end of the metal-free
star lifetime (a good approximation because of the short lifetime).
Therefore, the rate of increase of the entropy floor is
\begin{equation}
\dot K\fl(t)=\frac{\dot T\X(t)}{n_g^{2/3}(t)} =
 \frac{2\dot E_{reh}(t)}{3\kB n_g^{2/3}(t)} ~ ,
\label{dotK}
\end{equation}
where $\dot E_{reh}(t)$ is the rate of energy increase per particle,
\begin{equation}
\dot E_{reh}(t)=\dot N\SN(t) E\X \, /n_g ~,
\label{e:reh}
\end{equation}
and $\dot N\SN(t)$ is the rate of SNe per unit of volume, equal to the
star formation rate $\dot N_\star$ (eq.~[\ref{dndt}]) at the time
$t-t_{mf}(M_\star)$. The constant $E\X$ is the total energy per star
that is used to reheat the medium. Note that this is not the same as the
total energy radiated in X-rays by the star since some of that energy
goes into collisional ionizations or emission of recombination photons.
We leave the value of $E\X$ as a free parameter in our model (the range
of possible values is discussed in \S \ref{sres_xr}).

  Even though the reheating by X-rays is roughly homogeneous, the
increase of the entropy is of course lower in regions of high gas
density. To compute the entropy floor for a specific halo with formation
at $t_f$, we follow the evolution of the gas density during
the collapse assuming the spherical top-hat model, and we compute the
entropy increase from the previous history of the intensity of the
X-ray background generated by metal-free stars. The evolution of
$\dot N\SN(t)$ is computed self-consistently with the model from the
star formation rate. Heating by X-rays is no longer taken into account
after virialization (this would be negligible in any case once the gas
has reached a high density).

  X-rays have yet another effect: owing to secondary ionizations, they
can produce a higher ionized fraction than the residual value after
recombination (which is $3\times 10^{-4}$, see Galli \& Palla\ 1998).
This increases the formation rate of molecular hydrogen, enhancing the
cooling rate (Haiman, Rees, \& Loeb 1996). According to Shull \& van
Steenberg (1985), around 30 \% of the total energy in X-rays is used in
ionizations when the neutral fraction is high. The rate of ionizations
is then found by dividing expression (\ref{e:reh}) by 13.6 eV (the
ionization energy of hydrogen). Considering this effect, the evolution
of $f\fe$ is determined by
\begin{equation}
\dot f\fe(t)=-n_g k_1(T) f\fe(t)+0.3\dot E_{reh}(t)/13.6 eV,
\label{xr:fe}
\end{equation}
where $k_1(T)$ is the recombination rate, which can be approximated by
$k_1(T)=1.88\times 10^{-10} T^{-0.64} \cm^3$ s$^{-1}$ (Hutchins 1976) and
$n_g$ is the gas number density in the halo center after collapse. Before
collapse, the first term is negligible because the density is very low.
The initial value of $f\fe$ is given by the residual ionized fraction after
recombination.

  To summarize, for each halo mass at a fixed time, stars can form only
when the formation time satisfies equation (\ref{cool_c1}). When
considering feedback effects, cooling is delayed and therefore halo
formation has to take place earlier. Consequently, the probability that
a halo hosts a main sequence star in equation (\ref{prob_sh_star})
decreases, reflecting the feedback effects.

\subsection{Reionization}
\label{sreion}

  To calculate the reionization history of the IGM we must first
calculate the emissivity $\epsilon(t)$ of ionizing photons per IGM
baryon. We consider two contributions to the emissivity, one from
metal-free stars which we discuss in \S~\ref{semis_met_free}, and
another from stars with metals, discussed in \S~\ref{semis_met_rich}.
Finally, we explain the procedure to calculate the reionization history
in \S~\ref{semis_met_free}.

\subsubsection{Metal-Free Star UV Emission}
\label{semis_met_free}

  Metal-free stars emit a higher quantity of ionizing photons than
enriched stars (see Schaerer~(2002)). We use here the fitting formula
in Table 6 of Schaerer~(2002) to compute the ionizing luminosity of
each star as a function of its mass, $N\io(M_\star)$. For the values of
$M_\star$ we consider here, this yields $N\io(M_\star=100
\msun)=1.23 \times 10^{50}$ s$^{-1}$ and $N\io(M_\star=300 \msun)=4.52
\times 10^{50}$ s$^{-1}$.

 In addition, we take into account that some of the stellar photons may
be absorbed locally in the halo gas surrounding the star, and only a
fraction $f_{esc}$ of these photons will be able to escape and ionize
the IGM. We use the following simple model to compute $f_{esc}$
(see Whalen \etal 2004 for a more complete discussion and several
examples of the evolution of the \HII regions formed around a metal-free
star):
we assume that the star is initially surrounded by a distribution of
atomic gas following a singular isothermal profile. An \HII region grows
around the star, and the heated gas acquires a sound speed of $\sim 10
\kms$ and starts expanding. This speed is typically larger than the halo
circular velocity, but still smaller than the speed of the ionization
front as long as the escape fraction is not very small and the star is
capable of ionizing the halo (the stellar lifetime
is $\sim$ 3 million years, which is short compared to the crossing time
of the halo at this sound speed. Hence, the ionization front advances
supersonically. We use the simple approximation where all the gas moves
out by the same distance $r\cc(t) = 10\kms \times t$. The gas
distribution then has a central hole of radius $r\cc(t)$, and at $r>
r\cc(t)$ the density is
\begin{equation}
n\io(r,t)=n[r-r\cc(t),0]\frac{[r-r\cc(t)]^2}{r^2},
\end{equation}
where $n(r,0)$ is the initial gas number density profile of the halo.
The total rate of recombinations is computed from this density
profile assuming the halo gas has been fully ionized by the star, and
is subtracted from the photon emission rate of the star.

\begin{figure}
\plotone{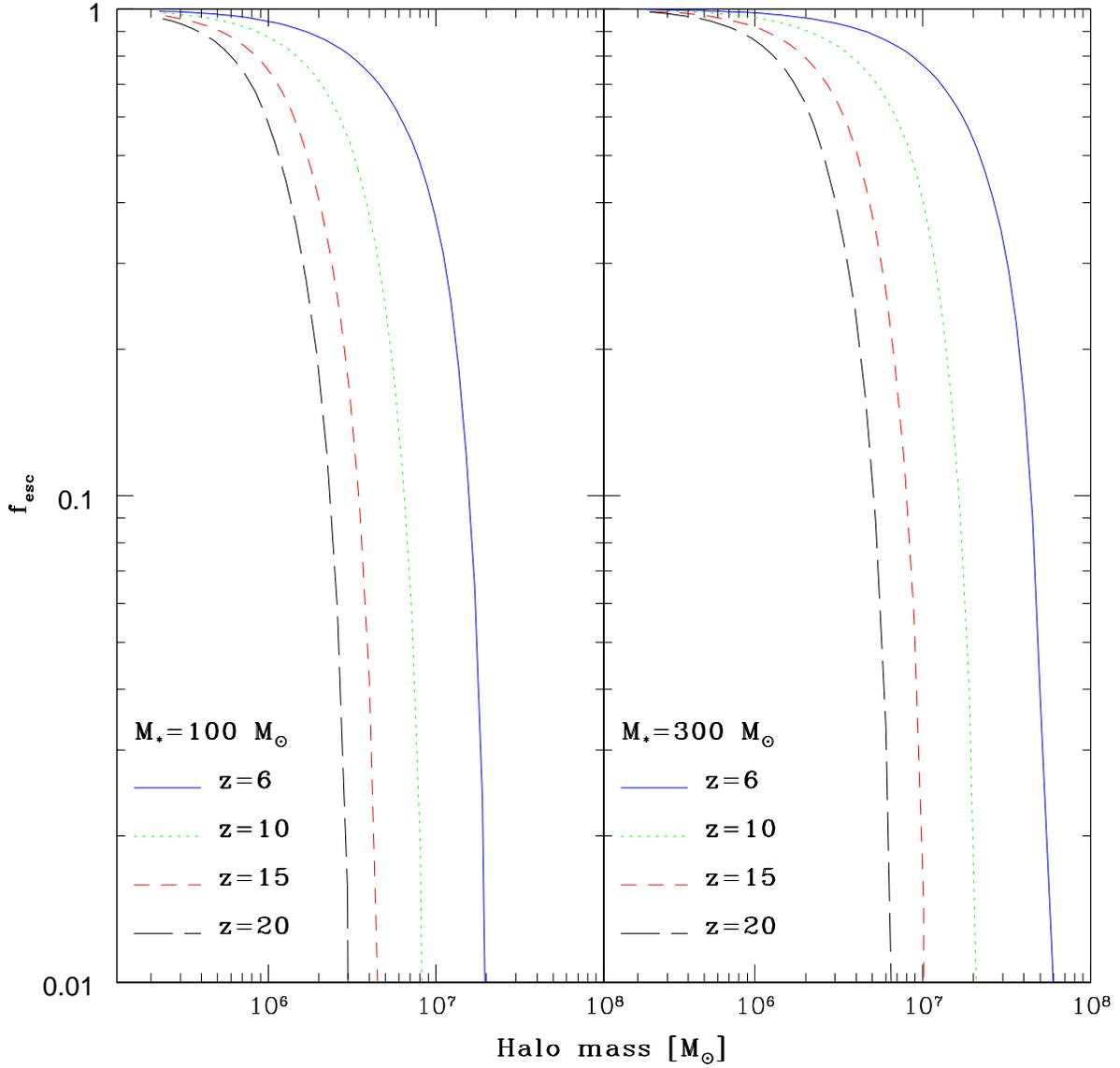}
\caption{ Escaping fraction of photons as a function of halo mass for
different redshifts, and a metal-free star with the two different masses
referred in each plot. Entropy is set as the primordial one.}
\label{ffesc}
\end{figure}

  The escape fraction of ionizing photons as a function of halo mass
($f_{\rm esc}(M)$) is plotted for different redshifts in Figure
\ref{ffesc}. We consider two possible masses of metal-free stars, $100$
\modot (left panel) and $300$ \modot (right panel). Naturally, the
escape fraction increases with stellar mass and decreases with halo
mass, because the number of recombinations in a halo with the gas fully
ionized increases with halo mass and does not depend on the luminosity
of the ionizing star. These results agree with the numerical simulations
of Kitayama et al.~(2004), suggesting that our simple model
is able to capture the main physics of the problem. The escape fraction
increases when reheating is considered, being more significant in low
mass halos, which have their density profiles flattened. Note that there
is a halo mass at which the escape fraction falls to zero; for halos
above this mass, the single central star is not able to ionize all the
halo gas and the \HII region reaches a Str\"omgren radius (the
assumption that all the halo gas is ionized is then obviously incorrect,
but this affects the calculation of $f_{esc}$ only when $f_{esc}$ is
already very small in any case).

  The global emissivity from metal-free stars, $\epsilon_{mf}$, is
finally calculated by integrating over all halo masses the number of
photons emitted per IGM particle corrected by absorption, i.e.,
\begin{equation}
\epsilon_{mf}(t)=\frac{1}{\bar{n}_g(t)}\int^\infty_0\der M\,N\h(M,t)
\, P_{ms}(M,t) \,f_{esc}(M)N\io(M_\star),
\label{eq:emis:mf}
\end{equation}
where $\bar n_g(t)$ is the baryonic number density of the IGM at $t$,
and $P_{ms}(M,t)$ is the probability that a halo harbors a main sequence
metal-free star (see eq.\ \ref{prob_sh_star}).

  For the emissivity of dissociating photons, we also use equation
(\ref{eq:emis:mf}) with $f_{esc}(M)=1$. The emission of dissociating
photons from metal-free stars, $N\diss(M_\star)$, is also computed from
the fit given in Table 6 in Schaerer~(2002) which yields
$N\diss(M_\star=100 \msun)=1.58\times 10^{50}$ s$^{-1}$ and
$N\diss(M_\star=300 \msun)=4.74 \times 10^{50}$ s$^{-1}$.

\subsubsection{Ionizing Emission from Second Generation Stars}
\label{semis_met_rich}

  After the formation of the first metal-free star in any halo, the
surrounding gas is polluted with the metals released by this star. The
metal-enriched gas can recombine and fall back into the halo. We assume
the metallicity of this infalling gas is then sufficiently high to
cool by the usual mechanisms that are prevalent in present galaxies
(Bromm et al.~2001). However, halos with metal-enriched gas below a
critical virial temperature may still not form stars because the cooling
rates are too small (e.g., Dijkstra et al.\ 2004b), and gas may remain
close to $\sim 10^4$ K in a spherical or thick disk distribution where
the gas is unable to cool further and fragment. In this paper, we adopt
the simple assumption that in halos were metal-rich star formation
takes place, the emissivity per unit baryonic mass is constant at all
redshifts. Then, the global emissivity due to enriched stars,
$\epsilon_{mr}(z)$, is calibrated to the observations of the ionizing
background intensity and mean free path of ionizing photons at redshift
$z=4$ (see Onken \& Miralda-Escud\'e 2004):
\begin{equation}
\epsilon_{mr}(z)=\epsilon_4 \frac{F(z)}{F_4},
\label{emisp12}
\end{equation}
where $\epsilon_4$ is the emissivity at redshift 4 in units of
ionising photons per particle and per Hubble time, $F(z)$ is the
fraction of mass in halos where metal-rich star formation takes place,
and $F_4$ is the same
fraction of mass at $z=4$. We assume that halos with circular velocity
above $35$ km/s are the ones forming stars at $z=4$ (as in Onken \&
Miralda-Escud\'e 2004; this affects only the normalization of the
emissivity), but consider different possibilities for the emissivity
at higher redshift, as explained below.

  In order to study the effect of varying the emission from normal,
metal-rich galaxies, we consider several models
listed in Table \ref{tcases}. These models are labeled by a number, and
in general any model will be referred to by the letter specifying the
feedback processes assumed (see Table 1) and this number. In Models 1
and 2, we compute $F(z)$ as the fraction of mass in metal-enriched
halos with a circular velocity greater than $10$ km/s, and in Models
3 to 6 we assume that all halos that have been polluted by metals form
enriched stars, with no minimum halo circular velocity. Note that even
by assuming a minimum halo circular velocity that is lower than at the
calibration redshift $z=4$, we are maximizing the emission at high
redshift under the constraint that the emission per unit mass is not
greater than in halos at $z=4$. The value of the normalization constant
$\epsilon_4$ is also subject to uncertainties arising from the process
by which the intensity of the ionizing background is inferred from
modelling the \lya forest transmitted flux, and the measurement of the
photon mean free path from the abundance of Lyman limit systems. We
adopt the values $\epsilon_4=7$ for Models 1 to 4 (as in
Miralda-Escud\'e~2003), and twice this value, $\epsilon_4=14$ (which
would be favored by the measurement of the ionizing background intensity
by Bolton et al.~2005), in Models 5 and 6.

  Finally, the emissivity from the metal-free stars is varied by
considering all stars to have a mass $M_\star = 100 \msun$ in models with
odd number, and $M_\star = 300 \msun$ in models with even number.

\subsubsection{Reionization History and Optical Depth}
\label{sreion_and_tau}

  The evolution of the ionized fraction $f\io(t)$ is calculated
according to
\begin{equation}
\frac{df\io(t)}{dt}=\epsilon(t)-\bar{n}_g^2(t)\, k_1\, f\io(t) ~,
\end{equation}
where $\epsilon(t)=\epsilon_{mf}(t)+\epsilon_{mr}(t)$ is the total
emissivity of ionizing photons, and $k_1$ is the recombination rate which
we evaluate at $T=10^4$ K. The cosmic history of $f\io(t)$
then yields the electron scattering optical depth $\tau_e$.

\section{Results}
\label{res}

  In this section, we present the results on metal-free star formation
and the reionization history of the IGM obtained from the approach
given in \S\ref{smethod}. In this model, there are four free
parameters: the stellar mass, $M_\star$, the reheating of IGM per
metal-free star, $E\X$, and the emissivity of metal-rich stars,
$\epsilon_4$.  In addition, there is some uncertainty in the physical
processes governing feedback effects. For this reason, we play with
the possibility to switch on or off the effect of photodissociating
photons in order to check how important it is. Finally, to compute
$F(z)$ in equation (\ref{emisp12}) we may consider or not halos with
circular velocity $< 10$ km/s (minihalos) to see whether the value of
$\tau\e$ experiences a big change. The models including all the
different possibilities are labelled XY, whith X running from A to D
and Y running from 1 to 6 (see Tables \ref{tcases} and
\ref{tsub_cases}).

  We now present the results on the rate of metal-free star formation
and the reionization history of the IGM for the models described in \S
2. The models and their nomenclature are described in Tables 1 and 2. The
model number (1 to 6) refers to properties of the radiation sources: the
mass of metal-free stars, and the normalization and low halo mass cutoff
for the emission from metal-enriched stars. The model letter (A to D)
refers to the presence or absence of the molecular photodissociation and
X-ray heating feedback effects. Results are first presented for the
case of no feedback effects (Models A, \S~\ref{sresultsnofeed}), then
including molecular photodissociation (Models B, \S~\ref{sresultsj21}),
and finally including X-ray photoheating (Models C and D,
\S~\ref{sres_xr}). Then, the reionization histories are reviewed
(\S~\ref{sresreion}). All models are run with the following set of
cosmological parameters: $\Omega\m=0.27$, $\Omega_{\Lambda}=0.73$,
$\Omega\ba=0.045$, $h_0=0.71$, $\sigma_8=0.84$ and $n=1$.

\begin{deluxetable}{cccc}
\tablecaption{{\it First column:} Source model number.
{\it Second column:} Emissivity normalization parameter for
metal-enriched stars, $\epsilon_4$. {\it Third column:}
Contribution of minihalos to metal-enriched stars.
{\it Fourth column:} Mass of metal-free stars. }

\tablewidth{0pt}
\tablehead{
\colhead{Source model} & \colhead{$\epsilon_4$} & \colhead{Minihalos} &
\colhead{$M_\star$ [\modot]}\\}
\startdata
1 & 7 & No & 100 \\
2 & 7 & No & 300 \\
3 & 7 & Yes & 100 \\
4 & 7 & Yes & 300 \\
5 & 14 & Yes & 100 \\
6 & 14 & Yes & 300 \\
\enddata
\label{tcases}
\end{deluxetable}

\begin{deluxetable}{ccc}
\tablecaption{{\it First column:} Feedback model letter.
{\it Second column:} Presence of a molecular photodissociating
background. {\it Third column:} Energy used to heat the intergalactic
medium by soft X-rays per metal-free star. }
\tablewidth{0pt}
\tablehead{
\colhead{Feedback model} & \colhead{Photodiss.} & \colhead{E$\X$ [erg]}\\}
\startdata
A & No & 0 \\
B & Yes & 0 \\
C & Yes & 10$^{51}$ \\
D & Yes & 10$^{52}$ \\
\enddata
\label{tsub_cases}
\end{deluxetable}

\subsection{No Feedback Effects}
\label{sresultsnofeed}

\begin{figure}
\plotone{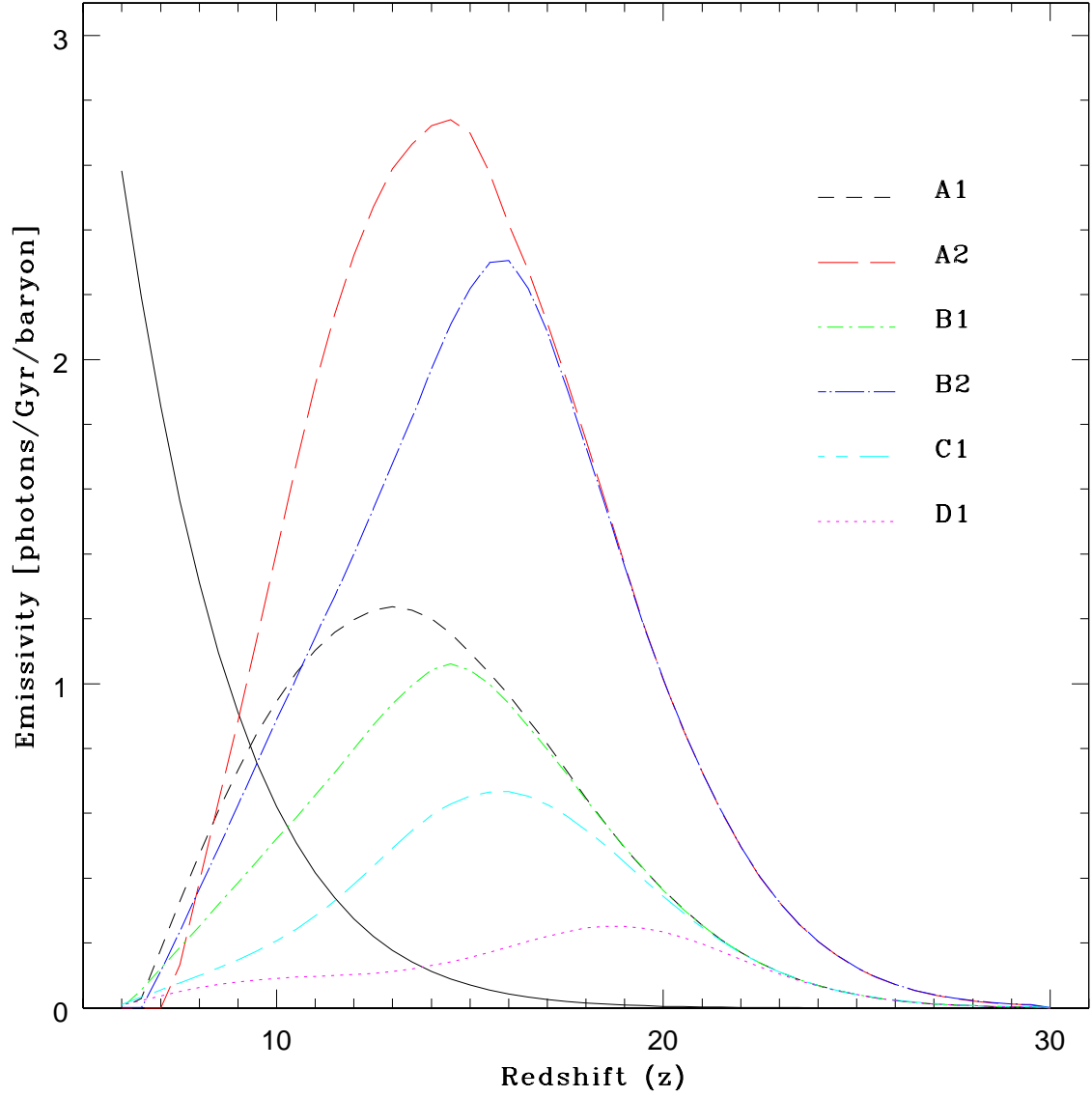}
\caption{ Emissivity as a function of redshift from metal-rich
(continuous line) and metal-free (dashed lines) stars for different
models. The total emissivity is the sum of the two.}
\label{femis}
\end{figure}

  The evolution of the emissivity for Model A1 is shown in Figure
\ref{femis}, both for metal-free and metal-rich stars. Metal-free stars
dominate at high redshift, and enriched stars start dominating below
$z\sim 10$. This behaviour is
easily understood. Initially there are only metal-free stars and their
contribution grows with time as more halos form. This results in an
increasing fraction of halos being polluted with metals, which at
some point causes a decline in the rate of new metal-free stars that
are formed. The halos collapsing with enriched gas start forming
enriched stars which eventually dominate the emissivity. The formation
of metal-free stars is also reduced as redshift decreases due to the
increased fraction of the IGM that is ionized. In ionized regions, the
gas is less able to cool and low-mass halos would continue merging into
more massive ones before forming any stars, implying a great reduction
in the number of metal-free stars that can be formed. This effect is
less important because the IGM ionized fraction remains relatively low
 at the epoch where most metal-free stars form.

  Comparing Models 1 (with metal-free stellar mass of $M_\star = 100
\msun$) with Models 2  ($M_\star = 300 \msun$), Figure 4 shows that the
emissivity of metal-free stars increases by a factor of $3$ as expected
at $z\gtrsim 20$, but at lower redshift it increases by a smaller factor
because the fraction of ionized IGM increases faster for $M_\star=300
\msun$, reducing the formation of metal free stars to a greater extent
than for Models 1. We note here that a high bias in the distribution
of metal-free stars relative to the mass could further increase the
effect of ionization in suppressing the number of metal-free stars that
can be formed.

\subsection{Effect of Molecular Photodissociation}
\label{sresultsj21}

  We now examine how the inclusion of a photodissociating background,
computed self-consistently from the star formation history, alters the
results. Figure 5 shows the evolution of the photodissociating
background intensity in Models B1 and B2. The maximum is reached
around $z\sim 15$ as the effects of increased metal pollution and
ionized IGM fraction suppress new metal-free star formation (note that
emission from enriched stars is not included in this figure). Figure 4
shows that the maximum of formation rate of metal-free stars occurs at
higher redshift when the photodissociating feedback is included, because
of the additional suppression of star formation. We also see in Figure
2 how the formation of metal-free stars is shifted to higher mass halos
in Models A compared to Models B at $z\lesssim 15$, because the negative
feedback delays the formation of the stars to a later stage at which
halos have grown to a higher mass.

\begin{figure}
\plotone{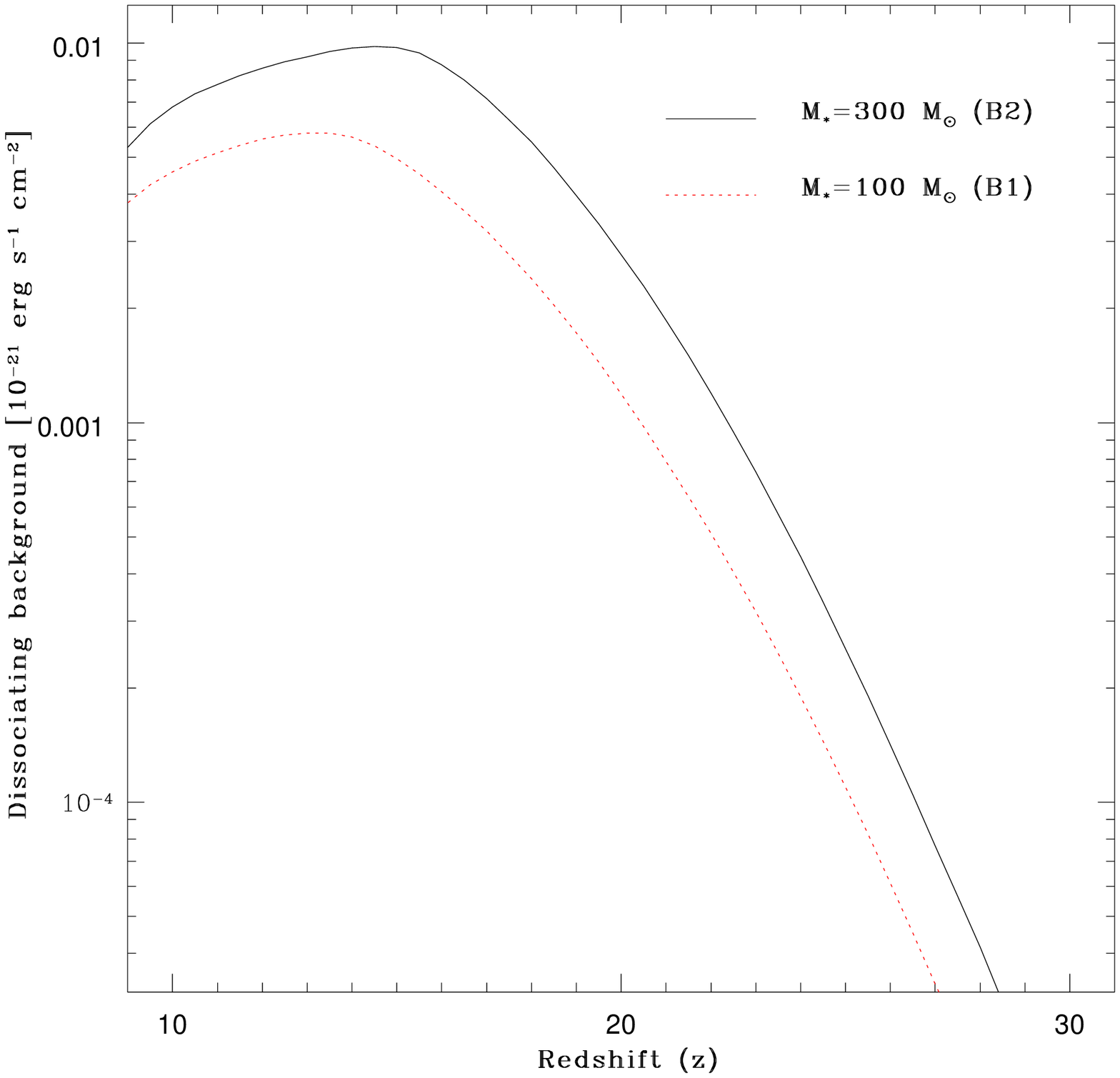}
\caption{ Photodissociating flux as a function of redshift for Models B1
and B2 (varying only the metal-free stellar mass)}
\label{fj21}
\end{figure}

\subsection{Effect of IGM Reheating by X-rays}
\label{sres_xr}

  We consider two models for the soft X-ray energy produced by a
metal-free star that is converted into IGM heat, $E_X$ (Table 2, Models
C and D). During the main-sequence, and at effective surface
temperatures near $\sim 10^5$ K, the fraction of radiative energy
emitted above $200 eV$ is $\sim$ 0.1\% of the total energy output of the
star, or $\sim 10^{51}$ erg. The supernova explosion at the end of the
life of a metal-free star can produce as much as $10^{53}$ erg of
kinetic energy for some stellar masses (see, e.g., Bromm et al.\ 2003),
and the fraction of these kinetic energy that is eventually reprocessed
into soft X-rays may be $\sim 0.03$ (Oh \& Haiman~2003), of which a
fraction of 50\% may be converted into IGM heat when the soft X-rays are
absorbed in a mostly neutral medium (Shull \& van Steenberg 1985). Given
the substantial degree of uncertainty in the soft X-ray energy emitted
per star, we consider the two values in Models C and D bracketing our
estimate.

  The increase in IGM temperature caused by the absorption of soft X-rays
is shown in Figure 6 for Models C1 and D1. The onset of formation of
metal-free stars causes a rise in temperature of the neutral medium
starting at $z\sim 25$, and at lower redshift the temperature ceases to
increase when the formation of metal-free stars is reduced (in practice
there would probably be an important contribution to the X-ray emission
from the enriched population, which has been considered here).

\begin{figure}
\plotone{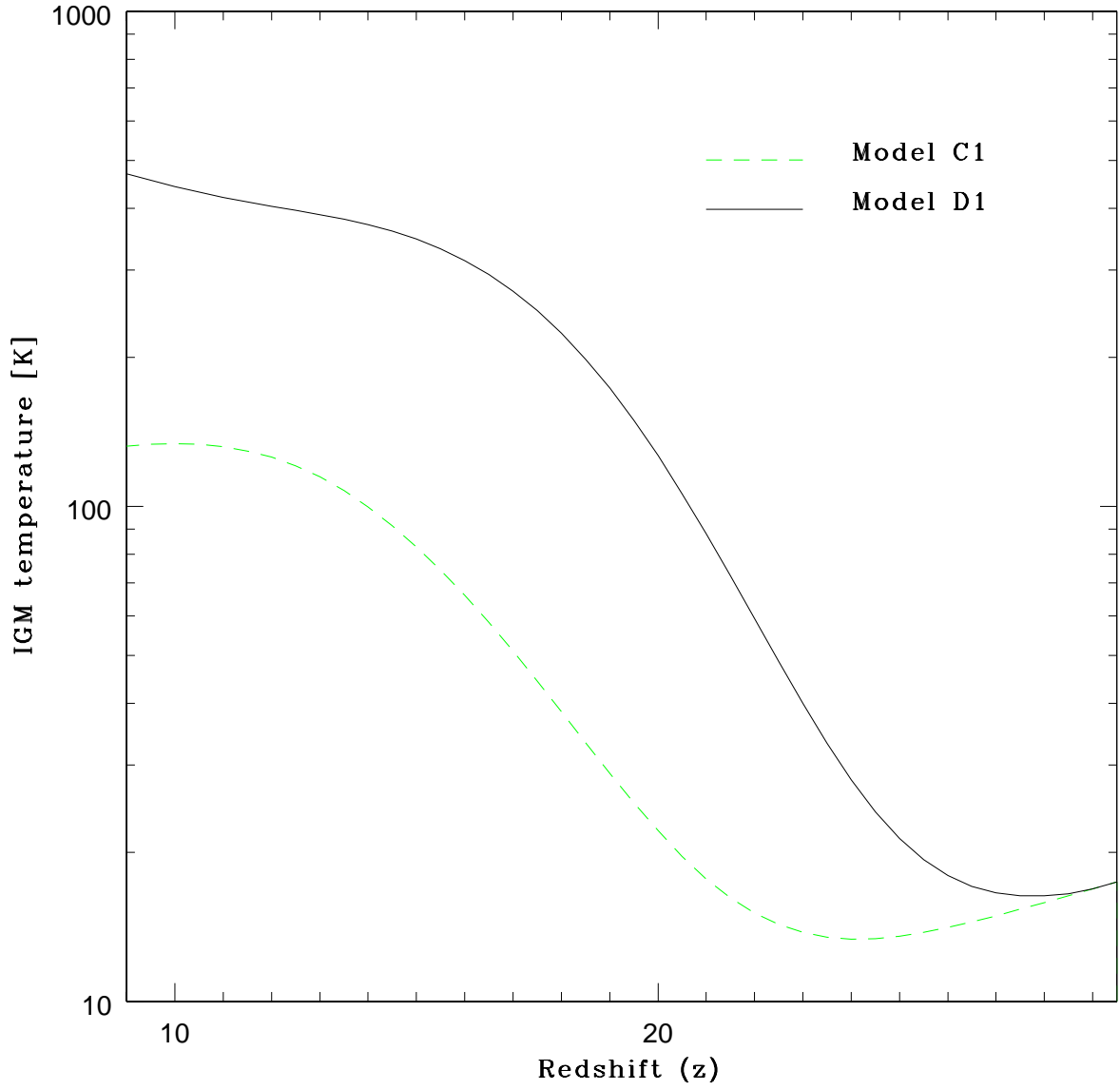}
\caption{ IGM temperature including soft X-ray heating. }
\label{ftemp}
\end{figure}

  The establishment of an entropy floor causes a large reduction of the
central density in low mass halos (see Fig.\ 1), which greatly reduces
the cooling rate. This implies that metal-free stars tend to be formed
in Models C and D in much more massive halos than in Models A and B: the
formation of these stars is suppressed in low-mass halos by the feedback
effect, so metal-free stars are not formed until later when a more
massive halo has formed from the mergers of several small ones (see
Figure 2). As long as our assumption that only one metal-free star forms
in each halo continues to hold, the metal-free stars will be much less
abundant. This is seen clearly in Figure 4, which shows the large
reduction in the emissivity of metal-free stars caused by the soft
X-ray heating.

\subsection{Reionization}
\label{sresreion}

\begin{deluxetable}{cccccc}
\tablecaption{Electron scattering optical depth to the CMB, $\tau\e$,
obtained for the different metal-free stellar masses and feedback
effects considered in this work.
\label{ttaue}}
\tablewidth{0pt}
\tablehead{ & & & & \multicolumn{2}{c}{X-rays ($E_X$)}\\
\cline{5-6}\\
\colhead{Model} & & \colhead{No feed.} & \colhead{Diss.} &
\colhead{$10^{51}$ erg} & \colhead{$10^{52}$ erg}\\
& & \colhead{A} & \colhead{B} & \colhead{C} & \colhead{D}\\}
\startdata
1 & \vline & 0.084 & 0.077 & 0.066 & 0.057\\
2 & \vline & 0.116 & 0.103 & 0.087 & 0.068\\
\hline
3 & \vline & 0.094 & 0.089 & 0.079 & 0.066\\
4 & \vline & 0.121 & 0.110 & 0.096 & 0.076\\
\hline
5 & \vline & 0.105 & 0.102 & 0.094 & 0.082\\
6 & \vline & 0.127 & 0.119 & 0.108 & 0.090\\
\enddata
\end{deluxetable}

\begin{figure}
\plotone{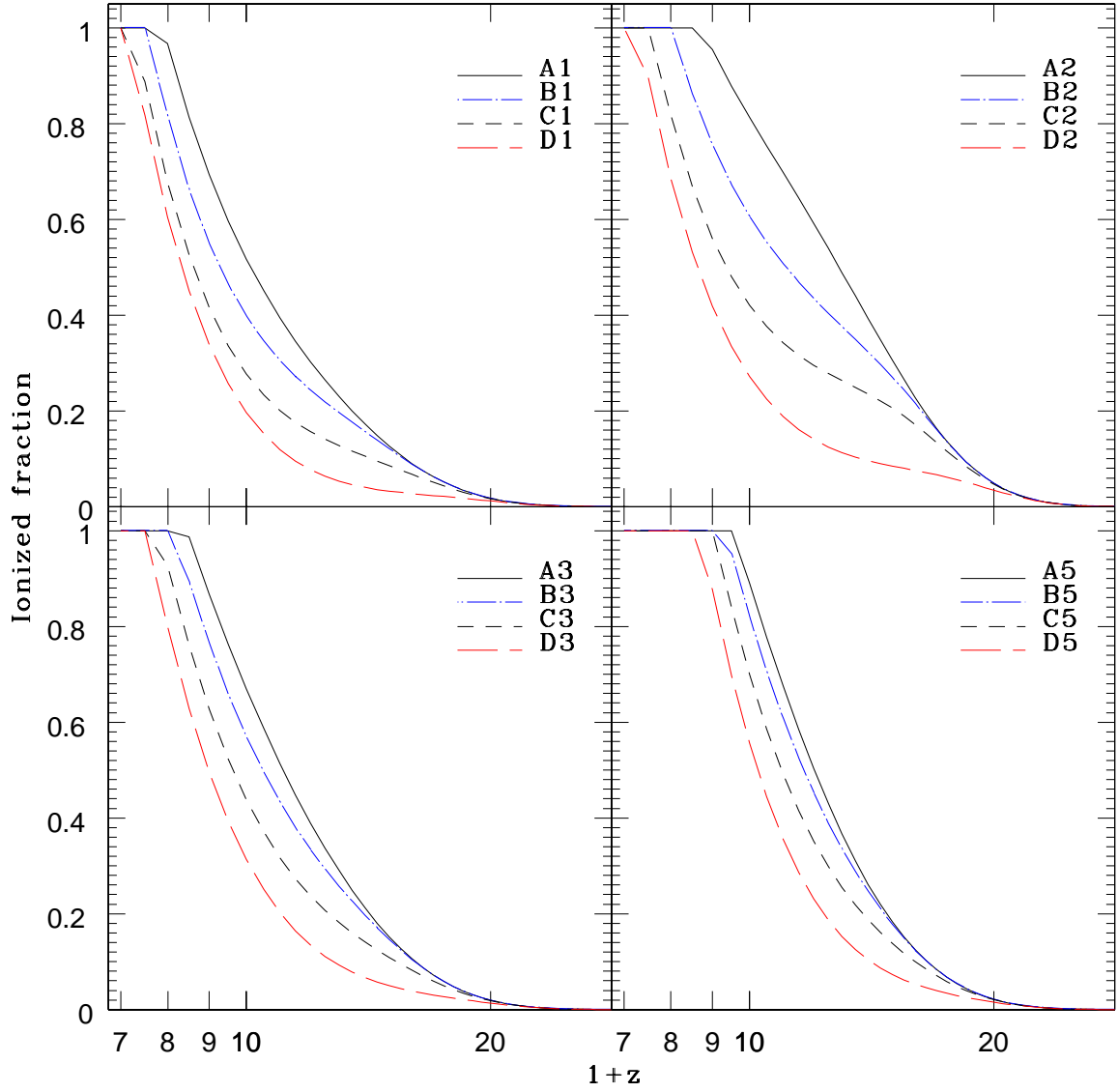}
\caption{ Ionized fraction as a function of redshift, for the indicated
models. }
\label{fionf}
\end{figure}

  Finally, we examine the reionization history and electron scattering
optical depth to the CMB of electron scattering in all the models.
In Figure \ref{fionf} we see that, as expected, feedback delays the
ionization of the IGM. Reionization always occurs gradually over a
relatively wide redshift range. Naturally, models with the highest
stellar mass and emission from enriched stars achieve a higher ionized
fraction at every redshift, and hence a higher optical depth. Even
though the total emissivity history may be double-peaked owing to the
presence of the metal-free and the enriched populations of stars (see
Fig.\ 4), there is never a double reionization because the recombination
rate is slow enough that the ionized fraction continues to increase when
the emission from metal-free stars starts decreasing.

  The redshift at which reionization is completed stays in the range of
$6-9$. This prediction of the redshift of the end of reionization is a
natural consequence of the value of the emissivity assumed for the
enriched stars that is derived from observations of \lya absorption 
systems at $z \lesssim 5$, and of assuming that the clumping factor of
ionized gas is not much larger than unity (Miralda-Escud\'e 2003). The
exact redshift of the end of reionization predicted by each model is
not an independent test of our models, because this redshift can be
modified and adjusted to agree with observations of the Gunn-Peterson
trough in the highest redshift quasars (White \etal 2003) by introducing
a moderate clumping factor with an adequate dependence on redshift (note
that we have assumed a clumping factor equal to one).

  The values of the electron scattering optical depth for each model
are presented in Table 3. The presence of metal-free stars has an
important effect in creating a small ionized fraction at very high
redshifts ($z\sim 20$). However, the increase of the optical depth that
metal-free stars can cause is modest and, even for $M_\star=300 \msun$
and in the absence of any negative feedback effects, they do not increase
the optical depth beyond $\sim 0.12$. Models D1, D3 and D5 represent
cases where the contribution from metal-free stars is minimal (due to
the very strong X-ray heating assumed; see Fig.\ 4). Comparing them
with models A2, A4 and A6 where the emission from metal-free stars is
maximum, we see that the early ionization caused by metal-free stars
can cause only a relatively modest increase of the optical depth,
under our central assumption that only one metal-free star is made
per halo, and no metal-free stars are made in halos that have already
merged with a halo in which star had previously formed.

%In order to better compare these results with observations we show in
%Table \ref{ttaue} the values of $\tau\e$ obtained for the different cases
%studied here. Clearly, values as high as the popular ones of WMAP cannot be
%achieved even in the more optimistic case. The most probable value seems to
%be around 0.1.

\section{Discussion and Conclusions}
\label{sdisc}

  We have considered in this paper the maximum emissivity and the
contribution to reionization that metal-free stars could make. Our
central assumption is that only one metal-free star is made in every
halo where pristine gas is able to cool via molecular hydrogen
rotovibrational lines. This is a reasonable assumption in view of the
simulations that have been made of the formation of the first star and
the effects of the subsequent ionizing radiation emitted and supernova
explosion: the gas cools hydrostatically towards the center, avoiding
fragmentation and forming a central star which can then ionize and
heat all the gas in its host halo, and probably expel it in the
supernova explosion (e.g., Abel \etal 2002; Bromm \etal 2002, 2003).
Once a star has exploded, it pollutes its host halo as well as every
larger halo into which its host halo merges in the future. Any halo
that has already formed a metal-free star in any of its halo
progenitors in the past is assumed to form a metal-enriched population
of stars that has an emissivity of ionizing photons similar to the
emissivity that is observationally determined at $z\sim 4$.

  Under this basic assumption, the contribution of the metal-free
stars to reionization is relatively modest. Metal-free
stars may dominate the emissivity at very high redshifts
($z\gtrsim 10$), but before they can manage to ionize much of the
universe, the metal pollution of halos rapidly reduces their formation
rate. Moreover, the effects of the ionization itself prevent cooling in
the remaining low-mass halos with pristine gas, delaying the formation
of a central metal-free star to the time of formation of more massive
halos that collapse later, and thereby reducing the number of metal-free
stars that are formed per unit of mass in the universe. Reionization was
likely completed by a population of stars and quasars in galaxies
forming from pre-enriched gas.

  We have analyzed the effects of molecular photodissociation and of
increasing the entropy floor by X-ray heating. These can further reduce
the amount of ionizing radiation that is emitted by the population of
metal-free stars. In agreement with Oh \& Haiman (2003), we have found
that the negative feedback induced by X-ray heating can in principle be
very large. However, as pointed out by Kuhlen \& Madau (2005), this
negative feedback might be much less important than suggested by the
simple calculation of Oh \& Haiman (2003) and the one we have presented
here. In our calculation, we consider only the cooling time of the gas
at the halo center in the hydrostatic equilibrium configuration obtained
at the end, by assuming an increased initial entropy produced by the
X-ray heating. In reality, the gas should be able to gradually cool and
lose entropy as it collapses in dense regions of halos, and these halos
later merge to form the final one where a star is formed. The same
X-rays that heat the gas would also increase the ionization, allowing
faster molecular cooling as the gas density increases, which might
greatly reduce the X-ray negative feedback effect. Nevertheless, our
main conclusion is that even without feedback effects the emission from
metal-free stars is strongly limited by the metal-pollution and the
ionization itself.

  Our models are generally in agreement with previous work. The value of
$\tau_e$ we infer is always lower than $0.13$, even when we increase
the emissivity of the enriched stellar population and we minimize any
negative feedback effects on the metal-free population. For a smaller
contribution from metal-free stars, values of $\tau_e$ are closer to
$0.10$ (see Table \ref{ttaue}), in agreement with the results of Ricotti
\& Ostriker (2004). Our results are also compatible with those of
Sokasian et al.~(2004),
%In particular, the reionized fraction at
%redshift 16 for
%our model B2 and that of model $M9_b$ in Sokasian et al.~(2004) sharing
%the same hypothesis are very similar (0.14 in our work and 0.12 in
%Sokasian et al.). However, we find stars in halos with masses lower
%than $10^6$ \modot contrary to Sokasian et al.~(2004). There are three
%possible reasons for this discrepancy.  First, we have used the
%extended Press-Schechter formalism to calculate the growth history of
%halos, while it is known that, for rare peaks, this formalism is not
%very accurate. A small deviation can produce different results
%particularly in the halo formation time distribution. Secondly, we
%have parameterized the NFW density profile with a concentration
%parameter $c=5$. However, this parameter changes with time and mass
%although its behaviour is not known at high redshifts. In particular,
%a lower value would produce a flatter density profile. Finally, we
%have considered a different prescription for the self-shielding. This
%is more restrictive than the one used by Sokasian et al.~(2004). We
%have checked what the results would be with no self-shielding, finding
%that the change is quite significant. In fact, the minimum halo mass
%forming stars is then $10^6$ \modot. In the rest of their models,
%Sokasian et al.
who find that high CMB optical depths to electron scattering can be
achieved only when many stars are assumed to be formed within halos
that cool by molecular hydrogen. The model M9 of Sokasian \etal is the
only one that assumes that only one star is formed per halo, and in this
case their results for the maximum optical depth are in good agreement
with ours. Similarly, Wise \& Abel (2005) can also produce
high optical depths only by assuming that many metal-free stars form in
each halo with a total mass much larger than in our model.

%These result are not surprising. If we consider that a metal-free star
%emits $10^5$ ionizing phtons per baryon, a 100 \modot metal-free star
%can ionize at most (if we neglect recombinations) $10^7$ \modot of
%neutral gas. This is much lower than the amount of neutral gas per
%comoving Mpc (around $10^{10}$ \modot). The number of halos per
%comoving Mpc that can form metal-free stars is fixed by gravitation
%(with correction for feedback mechanisms). Therefore, the only way to
%achieve significant ionized fractions at high redshift is by
%increasing the total mass of metal-free stars formed by halo.

  Naturally, if one is willing to assume a mass for metal-free stars
even larger than $300 \msun$, their contribution to the CMB optical
depth, $\tau_e$, can be further increased, but their total emission
increases by a factor smaller than the increase in the mass because of
the enhanced negative feedbacks. At the same time, if miniquasars
emitting copious amounts of ionizing radiation where produced by the
black holes created by the metal-free stars (which could occur when a
halo containing a central clump of cooling gas merges with another halo
that has already formed a star and contains a black hole in its center),
much more ionizing radiation could be produced, although the X-rays
emitted by these mini-quasars might then be in conflict with limits on
any unresolved component of the present soft X-ray background (Dijkstra
\etal 2004a).

  Our work can also be used to estimate the contribution of metal-free
stars to the Cosmic Infrared Background, which might include the
ultraviolet light of these stars from high redshift. The total number
of ionizing photons that are ever emitted by metal-free stars can be
computed straightforwardly by integrating the curves shown in Figure
4. The results are shown in Table \ref{temis} for each one of our
models. Even in the absence of negative feedbacks, and for stellar
masses $M_\star = 300\msun$, metal-free stars do not emit more than
$\sim$ one photon per baryon in the universe. This conclusion is not
surprising, since in the absence of many recombinations one does not
need a large number of ionizing photons to be emitted per baryon to
complete the reionization.  Because of the very high effective
temperatures of metal-free stars ($T\sim 10^5$ K), the number of
photons emitted in ultraviolet light at wavelengths longer than the
Lyman limit is similar. Hence, the contribution of metal-free stars to
the present Cosmic Infrared Background is not more than $\sim$ one
photon per baryon under the assumptions we have made.

  Recently, Kashlinsky \etal (2005) detected brightness fluctuations in
the Cosmic Infrared Background after subtracting all contributions from
known galaxies. The measured fluctuations are at a level of $\sim 0.1
{\rm nW}{\rm m}^{-2}\,{\rm sr}^{-1}$. This background intensity at a
wavelength of $5 \mu{\rm m}$ corresponds to $\sim 500$ photons per
baryon in the universe; as discussed by Kashlinsky \etal, in order to
account for the observed fluctuations the absolute brightness of the
infrared background contributed by metal-free stars would have to be as
high as $\sim 1 {\rm nW}{\rm m}^{-2}\,{\rm sr}^{-1}$ (or $\sim$ 5000
photons emitted per baryon), after taking into account the expected high
bias in the spatial distribution of these stars. It is clear that very
extreme assumptions about the number of metal-free stars that were
formed need to be made if the unaccounted fluctuations in the Cosmic
Infrared Background are related in any way to these metal-free stars. As
discussed by Santos \etal (2002), if metal-free stars are to make an
important contribution to the Cosmic Infrared Background, one needs to
assume that a large fraction of all the gas in each halo where pristine
gas cools by molecular hydrogen (typically more than $10^5 \msun$) forms
metal-free stars, and that most of the radiation from the stars is
somehow internally absorbed in the halos in order to prevent an
excessively early reionization of the IGM. Both of these requirements
are not realistic: once a central star is formed in a $\sim 10^6\msun$
halo, its ionizing radiation will ionize and push all the halo gas out
over the short main-sequence lifetime of the star, and the escape
fraction of the ionizing photons is high except in rare cases where a
single star forms in a very massive halo (see Figure \ref{ffesc}). The
detected Cosmic Infrared Background fluctuations have other more likely
possible sources, such as an incomplete accounting of the faint-end of
the galaxy luminosity function, or a normal population of low-luminosity
galaxies with metal-enriched gas and stars at high redshift.

\acknowledgments
  JM acknowledges helpful conversations with Mark Kuhnen, Piero Madau,
and Daniel Wang. JM would also like to thank the Institute for Advanced
Study for their hospitality, where part of this work was completed.
This work was supported in part by the Direcci\'on General de
Investigaci\'on Cient\'\i fica y T\'ecnica of Spain, under contract
AYA2003-07468-C03-01. JMR was supported by a fellowship of the
Ministerio de Educaci\'on, Cultura y Deporte of Spain.

\appendix

\section{Star Formation in Haloes}

%Specifically, what we need is the probability for a given halo with
%$M$ at $t$ to form a star right at that time. This probability arises
%from two factors: the probability that cooling has operated enough
%time since the formation of the halo for a metal-free star to be able
%to form in it, and the probability that no metal-free star has formed
%in any ancestror of the halo. Next we show how to calculate the second
%of these two (tightly related) quantities.

  Consider a halo of mass $M$ at time $t$. The average number of
smaller halos with mass between $M'$ and $M'+dM'$ at $t' < t$ that are
located within the halo of mass $M$ at $t$ is given by the extended PS
formalism,
\begin{equation}
N_{LC}(M',t'\rightarrow M,t)dM' = \frac{M}{M'}\,
\frac{df\left(M',t' \vert M,t\right)}{dM'}dM',
\label{dfdm}
\end{equation}
where $df/dM'$ is the conditional probability for a mass element to be
part of a halo of mass $M'$ at $t'$, given that it is part of a larger
halo of mass $M>M'$ at a later time $t>t'$ (eq.\ [2.16] in Lacey \& Cole
1994). The rate at which stars form at $t'$ in halos of
mass between $M'$ and $M'+dM'$, which will be found within the halo of
mass $M$ at $t$, is
\begin{equation}
{dN_\star(M',t'\rightarrow M,t) \over dt'}dM' =
 N_{LC}(M',t'\rightarrow M,t)dM' \frac{dP_\star(M',t')}{dt'},
\label{nstar}
\end{equation}
where $dP_\star(M',t')/dt'$ is the probability per unit of time that a
halo with mass $M'$ forms a star at $t'$.

This latter probability is obtained by noting that the probability for
a star to form in a halo with mass $M'$ between $t'$ and $t'+dt'$ is
equal to the probability for the corresponding halo to form between
$t\f$ and $t\f + dt\f$, being $t\f(M',t')-t'$ the required cooling time
in this halo to form a star at $t'$, times the probability
$P_\star\first(M',t')$ that no halo ancestor has previoulsy formed
any star,
\begin{equation}
\frac{d P_\star(M',t')}{d t'}\,d t'=P_\star\first(M',t')\,
\frac{d P(M',t',t\f)}{ dt_f}\, dt_f ~.
\label{i}
\end{equation}
In equation (\ref{i}), $dP(M',t',t_f)/dt_f$ is the probability distribution
function of halo formation times $t_f$, and the derivative $dt\f/dt'$ is to be
evaluated by requiring the condition $t'=t\f + t_{cool}(M',t')$.
Although a better behaved definition of halo formation time with known
analytical expressions of the corresponding probability distribution
function can be found in the literature (Raig et al.\ 2001), for
simplicity we use here that provided by Lacey \& Cole (1993). These
authors define the formation time of a halo with mass $M$ at $t$ as the
earliest time $t\f<t$ at which some progenitor reaches a mass
$M'\ge M/2$.

  At the same time, the average number of stars that have previously
formed in ancestors of the halo with mass $M$ at $t$, $N_\star(M,t)$, is
obtained by integration of equation (\ref{nstar}),
\begin{equation}
N_\star(M,t)=\int_{0}^t dt'\int^M_0 \der M' N_{LC}(M',t'\rightarrow M,t)
\frac{d P_\star}{d t'}[M',t']\,.
\label{nave}
\end{equation}
Assuming a Poisson distribution of the number of metal-free stars formed
in ancestors of a halo of mass $M$ at $t$, with mean expectation value
provided by equation (\ref{nave}), the probability that no ancestor
has previously formed any metal-free star is
\begin{equation}
P_\star\first(M,t)=e^{-N_\star(M,t)}\,.
\end{equation}
Notice that in order to calculate the previous probability we must
proceed in an iterative way since the term $d P_\star/d t'$ in
equation (\ref{nave}) includes the factor $P_\star\first(M',t')$
affecting its ancestors, implying that we need to save at each step the
information of all previous halos along the merger tree.

  Another useful quantity is the metal-free cosmic star formation rate,
$d N_\star/d t$. This is readily obtained from $dp_\star/dt$
(eq.\ [\ref{i}]) by integrating over all masses with the halo mass
function $N\h(M,t)$

\begin{equation}
\dot N_\star(t)=\int^\infty_0 d M\,N\h(M,t)
\,\frac{d P_\star[M,t]}{d t}\,.
\end{equation}

\begin{deluxetable}{cccccc}
\tablecaption{Total number of photons emitted per baryon by metal-free
stars obtained for the different metal-free stellar masses and feedback
effects considered in this work.\label{temis}}
\tablewidth{0pt}
\tablehead{ & & & & \multicolumn{2}{c}{X-rays ($E_X$)}\\
\cline{5-6}\\
\colhead{Model} & & \colhead{No feed.} & \colhead{Diss.} &
\colhead{$10^{51}$ erg} & \colhead{$10^{52}$ erg}\\
& & \colhead{A} & \colhead{B} & \colhead{C} & \colhead{D}\\}
\startdata
1 & \vline & 0.46 & 0.33 & 0.18 & 0.07 \\
2 & \vline & 0.80 & 0.61 & 0.40 & 0.19 \\
\hline
3 & \vline & 0.39 & 0.29 & 0.16 & 0.06 \\
4 & \vline & 0.73 & 0.56 & 0.38 & 0.18 \\
\hline
5 & \vline & 0.31 & 0.24 & 0.14 & 0.06 \\
6 & \vline & 0.62 & 0.49 & 0.34 & 0.16 \\
\enddata
\end{deluxetable}

\end{document}